# Generalized parallel concatenated block codes based on BCH and RS codes: construction and Iterative decoding

A. Farchane, M. Belkasmi and S. Nouh

**Abstract**— In this paper, a generalization of parallel concatenated block (GPCB) codes based on BCH and RS codes is presented. On the sender side two systematic encoders are used and separated by an interleaver. In the receiver side, the Chase-Pyndiah algorithm is utilized as component decoder in an iterative process. The effect of various component codes, interleaver sizes and patterns, and the number of iterations are investigated using simulations. The simulation results show that the slope of curves and coding gain are improved by increasing the number of iterations and/or the interleaver size. From the simulations we observe that the performance GPCB based on RS become worse when increasing the length of the component codes. This is differing with the knowledge that increasing the codes' length leads to better performance.The comparison between GPCB based on BCH only, RS only or BCH and RS codes shows that the performance of the GPCB based on BCH codes are the best and the ones based on BCH and RS codes fall in between the two other GPCB codes.

## 1 Introduction

The exceptional performance of turbo codes was first demonstrated by C. Berrou [1] in 1993. To construct the turbo code they used concatenated recursive convolutional codes with a non-uniform interleaver inserted between two recursive convolutional encoders. In 1994, to achieve near optimum decoding of product codes R. Pyndiah [2] proposed a new iterative decoding algorithm based on a soft output decision version of Chase decoding. The obtained results are similar to convolutional turbo codes and the coding gains were close to the theoretical coding gain expected from product codes when using maximum likelihood sequence decoding [6].

It seems that the construction of GPCB codes was introduced independently by Nilsson et al [7], and Benedetto et al [5]. Nilsson et al. used a MAP based iterative decoding and evaluated the new scheme by simulations. Whereas Benedetto et al evaluate theoretically the performances of GPCB codes. Also the construction given by Nilsson is more general than that of Benedetto.

The generalized parallel concatenated block (GPCB) codes are similar to convolutional turbo codes in encoding and decoding structure. Iterative decoding of concatenated codes is a way of using long powerful codes while keeping the decoder relatively simple. A time and safety critical application benefits from the long powerful codes. In [8], we consider generalized parallel concatenated block (GPCB) codes based on the BCH family. In this work, we demonstrated the application of the Chase-Pyndiah SISO algorithm to decode the GPCB-BCH codes, and investigated the effects of various parameters on the performance using simulations. An extension of this work is to study the performances of the generalized parallel concatenated RS (GPCB-RS) codes, and the generalized parallel concatenation based on the two families. It means BCH and RS codes. We denote by GPCB-BCH-RS the parallel concatenation of BCH and RS codes.

Our study is based on RS and BCH codes, we decode by using a Chase-Pyndiah algorithm. The effects of various component codes, the number of iterations, interleaver size and pattern are investigated using simulations.

In this paper, the first section describes the encoder structure of the generalized parallel concatenated block codes. Section III presents the component decoder. Section IV describes the iterative decoding of the GPCB codes. The simulation results are given in section V. In section VI we compare the performances of the GPCB-BCH codes with those of the GPCB-RS and GPCB-BCH-RS codes. Section VII concludes this paper.

*A. Farchane[1], M.Belkasmi[2] and S.Nouh[3] are with the Department of Communication Technology, Mohammed V Souissi University, Rabat, National School of Computer Science and Systems Analysis (ENSIAS), 10000, Morocco. E-mails:*
*farchane@yahoo.fr[1], Belkasmi@ensias.ma[2], nouh_ensias@yahoo.fr[3]*

## 2 Generalized parallel concatenated block codes

The structure of the GPCB encoder is shown in figure 1. Two systematic block encoders are used as component codes with an interleaver placed before the second block encoder. We have explored in this study two constructions of GPCB codes.

### 2.1 Construction 1

Here a block of N = M.k data symbols at the input of the encoder is subdivided to M sub-blocks each of k symbols. Each k symbols vector is encoded in order to produce n symbols codeword. The input block is scrambled by the interleaver-denoted by Π- before entering in the second encoder. The codeword of GPCB code consists of the input block followed by the parity check symbols of both encoders. In this contribution, several interleaving techniques were invoked such as random, block, diagonal, cyclic and Berrou's interleaver [9], [10].

TABLE 1: SOME EXAMPLES OF GPCB CODES

| Component code 1 | Component code 2 | M | GPCB code | Rate(GPCB code) |
|---|---|---|---|---|
| BCH(63, 51, 5) | BCH (63, 51, 5) | 1 | GPCB-BCH (75, 51) | 0.68 |
| | | 10 | GPCB-BCH (750, 510) | 0.68 |
| | | 100 | GPCB-BCH (7500, 5100) | 0.68 |
| | | 1000 | GPCB-BCH (75000, 51000) | 0.68 |
| BCH(127, 113, 5) | BCH (127, 113, 5) | 1 | GPCB-BCH (141, 113) | 0.80 |
| | | 10 | GPCB-BCH (1410, 1130) | 0.80 |
| | | 100 | GPCB-BCH (14100, 11300) | 0.80 |
| | | 1000 | GPCB-BCH (141000, 113000) | 0.80 |
| BCH(255, 239, 5) | BCH (255, 239, 5) | 1 | GPCB-BCH (271, 239) | 0.88 |
| | | 10 | GPCB-BCH (2710, 2390) | 0.88 |
| | | 100 | GPCB-BCH (27100, 23900) | 0.88 |
| | | 1000 | GPCB-BCH (271000, 239000) | 0.88 |
| RS(63, 53, 11) | RS(63, 53, 11) | 1 | GPCB-RS(73, 53) | 0.72 |
| | | 10 | GPCB-RS(730, 530) | 0.72 |
| | | 100 | GPCB-RS(7300, 5300) | 0.72 |
| | | 1000 | GPCB-RS(73000, 53000) | 0.72 |
| RS(127, 115, 13) | RS(127, 115, 13) | 1 | GPCB-RS(139, 115) | 0.82 |
| | | 10 | GPCB-RS(1390, 1150) | 0.82 |
| | | 100 | GPCB-RS(13900, 11500) | 0.82 |
| | | 1000 | GPCB-RS(139000, 115000) | 0.82 |
| RS(255, 243, 13) | RS(255, 243, 13) | 1 | GPCB-RS(267, 243) | 0.91 |
| | | 10 | GPCB-RS(2670, 2430) | 0.91 |
| | | 100 | GPCB-RS(26700, 24300) | 0.91 |
| | | 1000 | GPCB-RS(267000, 243000) | 0.91 |
| BCH (63, 51, 5) | RS(63, 51, 13) | 1 | GPCB-BCH-RS(75, 51) | 0.68 |
| | | 10 | GPCB-BCH-RS(750, 510) | 0.68 |
| | | 100 | GPCB-BCH-RS(7500, 5100) | 0.68 |
| | | 1000 | GPCB-BCH-RS(75000, 51000) | 0.68 |
| BCH (127, 113, 5) | RS(127, 113, 15) | 1 | GPCB-BCH-RS(141, 113) | 0.80 |
| | | 10 | GPCB-BCH-RS(1410, 1130) | 0.80 |
| | | 100 | GPCB-BCH-RS(14100, 11300) | 0.80 |
| | | 1000 | GPCB-BCH-RS(141000, 113000) | 0.80 |
| BCH (255, 239, 5) | RS(255, 239, 17) | 1 | GPCB-BCH-RS(271, 239) | 0.88 |
| | | 10 | GPCB-BCH-RS(2710, 2390) | 0.88 |
| | | 100 | GPCB-BCH-RS(27100, 23900) | 0.88 |
| | | 1000 | GPCB-BCH-RS(271000, 239000) | 0.88 |

A systematic GPCB code is based on two component systematic block codes, $C_1$ with parameters ($n_1$, k) and $C_2$ with parameters ($n_2$, k). Viewing the coding scheme of figure 1 as single GPCB encoder, the length of the information-word to be encoded by the GPCB code is given by the size of the interleaver N=M.k. The first encoder produces $P_1 = M \times (n_1 - k)$ parity check symbols. The second encoder produces $P_2 = M \times (n_2 - k)$ parity check symbols. Thus the total number of parity symbols generated by the GPCB encoder is: $P = P_1 + P_2 = M \times (n_1 + n_2 - 2 \times k)$. The length of the GPCB codeword is given by: $L = N + P = M \times (n_1 + n_2 - k)$ Consequently, the code rate of the GPCB codes can be computed by: $\Re = N / L = k / (n_1 + n_2 - k)$.
This implies that the GPCB code rate is independent of the interleaver size N. Table 1 gives some examples of codes based on this construction.

### 2.2 Construction 2

Generalized parallel concatenation can also be done using two or more codes belonging to two different families (e.g. BCH and RS codes). Constituent codes must have the same length. Let us consider two codes BCH(n, $k_1$) et RS(n, $k_2$), where $k_2 = n - 2t_2$ and $k_1 = n - mt_1$, where t1 and t2 is the error correction capabilitie of the BCH and RS codes respectively and m is equal to log2(n+1). The equality $k_1 = k_2 = k$ imply that $m.t_1 = 2t_2$, it mean that $t_2 = \frac{m.t_1}{2}$. If m or t1 is even, then we can find two codes RS and BCH with the same dimension.

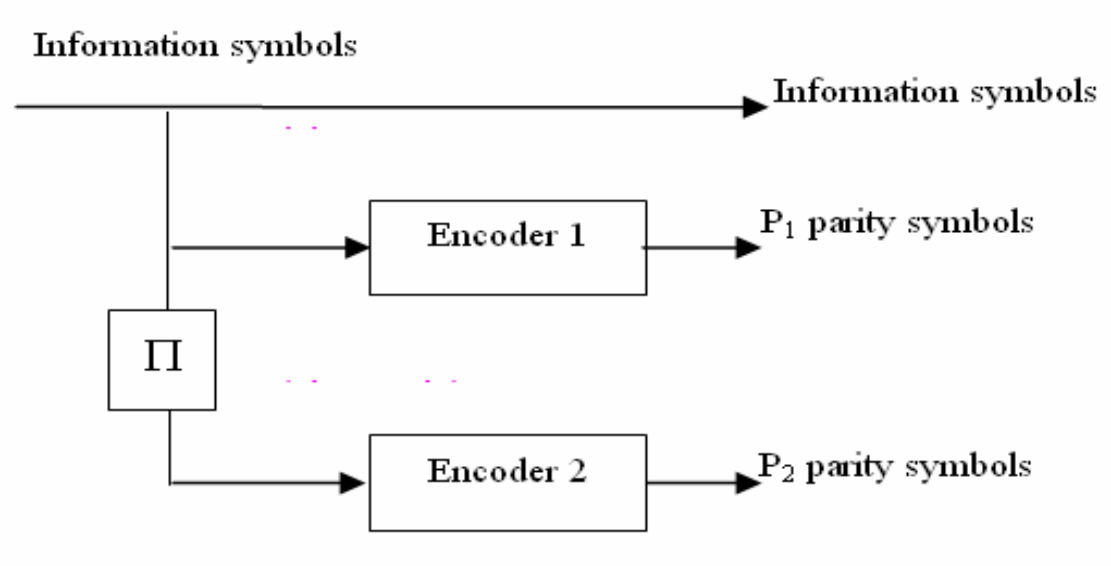

Figure 1: Encoder structure of parallel concatenated block (GPCB) codes

A block of $N = M.m.k$ bits at the input of the encoder is subdivided to $M.m$ sub-blocks each of k bits. Each k bits vector is encoded in order to produce n bits codeword. The input block is scrambled by the interleaver, denoted by Π. Each m bits are replaced by the corresponding symbol in the field GF(2m). The later operation transforms a block of $M.m.k$ bits to a block of $M.k$ symbols. The block of symbols is subdivided into M sub-blocks of length k before entering the second encoder.
The codeword of GPCB code consists of the input block followed by the parity check of both encoders. Table 1 gives some examples of codes based on this construction.

In the same maner two RS codes with length n form $2^m$-1 and dimension k can be used to construct of a GPCB-RS code.

## 3 COMPONENT DECODER

We consider a transmission that use BPSK modulation coded by a block code, with code rate $k_i/n_i$ (i=1 or 2).
The input of the decoder, when the channel is perturbed by a white Gaussian noise, is equal to $R = C + B$, where $R = (r_1 \ldots r_j \ldots r_{n_i})$ is the observed vector, $C = (c_1 \ldots c_j \ldots c_{n_i})$ $c_j = \pm 1$ is the transmetted codeword and $B = (b_1 \ldots b_j \ldots b_{n_i})$ is the white noise whose components $b_j$ have zero average and variance $\sigma^2$.
We choose as component decoder the Chase-Pyndiah algorithm [11]. This decoder works as follows:
The decoder starts by generating a set of codewords which are in the vicinity of the received vector R. Then, among those codewords, it selects the nearest codeword from R in term of Euclidean distance. By doing that it tries to determine the most likelihood codeword. The reliability of the decoded bits is given by the log likelihood ratio (LLR) of the decision $d_{if}$ which is defined by:

$$\mathrm{LLR}_{jf} = \ln \frac{\Pr(e_{jf} = +1/R)}{\Pr(e_{jf} = -1/R)} \qquad (1)$$

Where $e_{jf}$ is the binary element in position (j, f) of the transmitted code word E, $1 \le j \le n$ and $1 \le f \le m$.
The expression of the $LLR_{jf}$ can be approximated, in the case of the AWGN, by:

$$LLR_{jf} = \frac{1}{2\sigma^2}\left[\left|R - C^{\min(-1)}\right|^2 - \left|R - C^{\min(+1)}\right|^2\right] \qquad (2)$$

Where $C_{jf}^{\min(+1)}$ and $C_{jf}^{\min(-1)}$ are two codewords at minimum Euclidean distance from R with $C_{jf}^{\min(+1)} = +1$ *and* $C_{jf}^{\min(-1)} = -1$ , $C_{jf}^{\min(+1)}$ *and* $C_{jf}^{\min(-1)}$ are chosen among the subset of code word given by Chase algorithm. By expanding relation (2) we obtain:

$$LLR_{jf} = \frac{2}{\sigma^2}\left( r_{jf} + \sum_{\substack{x=1 \\ x\neq j}}^{n} \sum_{\substack{z=1 \\ z\neq f}}^{m} r_{xz} c_{xz}^{\min(+1)} \rho_{xz} \right)$$

Where $((x,z) \neq (j,f))\ \rho_{xz} = \begin{cases} 0\ if\ C_{xz}^{\min(+1)} = C_{xz}^{\min(-1)} \\ 1\ if\ C_{xz}^{\min(+1)} \neq C_{xz}^{\min(-1)} \end{cases}$

If we normalize the approximated LLR of $d_{if}$ with respect to $2/\sigma^2$ we obtain:

$$r'_{jf} = (\sigma^2/2).LLR_{jf} = r_{jf} + w_{jf}$$

The estimated normalized LLR of of decsion $\boldsymbol{d_{if}}$, $r'_{jf}$ is given by input samples $r_{jf}$ plus $w_{jf}$ which is independent of $r_{jf}$. The LLR of $r'_{jf}$ is an estimation of the soft decision of the RS decoder.
To compute the normalized $LLR_{jf}$ of binary elements at the output RS decoder, we must first select the codeword at minimum Euclidean distancefrom R. Let $C^{\min(+i)}$ be this code word, $C^{\min(+i)}$ has a binary element I at position $(j,f)$ (i=±1). Then we look for codeword Cmin(+i) at minimal Euclidean distance from R among the codeword subset obtained by Chase algorithm.

$C^{\min(-i)}$ must have -i as binary element as position (j,f).

If the $C^{\min(-i)}$ codeword is found, the soft decision $r'_{jf}$ of $\boldsymbol{d_{if}}$ can be computed using the relation given bellow:

$$r'_{jf} = \left( (M^{\min(-i)} - M^{\min(i)}) / 4 \right) c_{jf}^{\min(i)}$$

Where $M^{\min(-i)}$ and $M^{\min(i)}$ represent respectively the $C_{jf}^{\min(-1)}$ Euclidean distance from R and $C_{jf}^{\min(+i)}$ Euclidean distance from R.
Else we use the relation: $r'_{jf} = \beta.c_{jf}^{\min(i)}$ where β is a constant which is a function of the iteration.

# 4 ITERATIVE DECODING OF GPCB CODES

## 4.1 GPCB decoder

The decoding of the GPCB codes is iterative. The decoder structure is shown in figure 2. An iteration consists in using two component decoders serially. The first one uses the systematic information and the first parity check symbols in order to generate extrinsic information as in the Chase-Pyndiah algorithm. This extrinsic information is used to update the reliabilities of the systematic information which will be interleaved and feed into the second decoder with the second parity check symbols received from the channel. The second decoder also generates the extrinsic information using Chase-Pyndiah decoder, and then updates the reliabilities of the systematic information for the second time. The updated reliabilities will be desinterleaved and feed again into first decoder, for the next iteration. The process resume until a maximum number of iterations is reached. The coefficients α and β used in Chase-Pyndiah algorithm are listed in table 2.

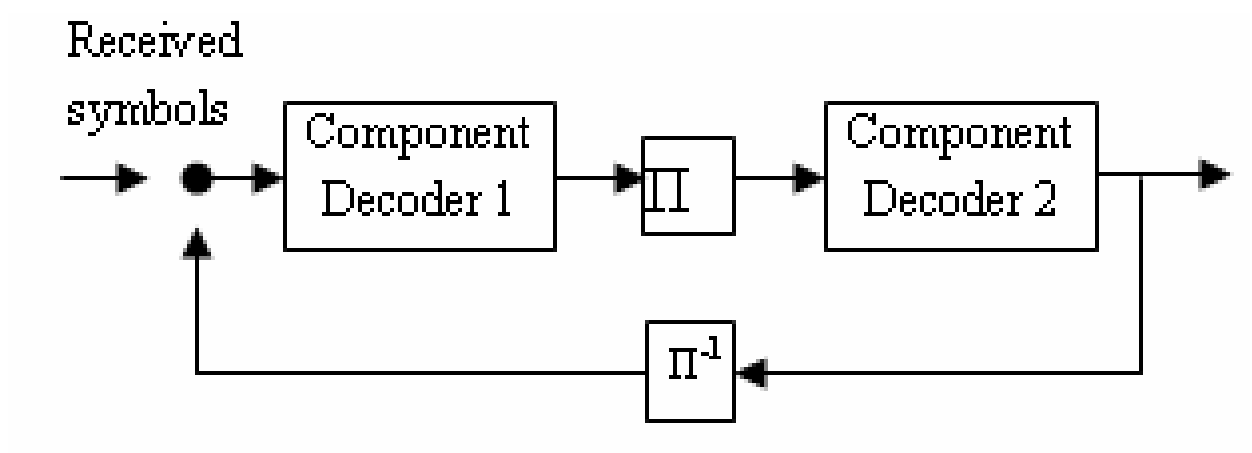

Figure 2: Iterative decoding structure for the GPCB codes

## 4.2 Parameters $\alpha$ and $\beta$

We have determined the values of $\alpha$ and $\beta$ empirically. The later parameters play a crucial role to have good performance. So the better parameters you have the better performance you will gain. Therefore, we should carefully determine these parameters. To obtain good parameters, we choose some condition for which codes are sensitive. Thus we take the parameter M equal to 100, and relatively high component code length.

We begin our process by setting the number of iterations in 1, and vary the value of $\alpha$, where $0 \leq \alpha \leq 1$, in order to have good performance, and keep the value of $\alpha$ which gives the best BER (bit error rate). Next, we vary the value of the parameter $\beta$, where $0 < \beta \leq 1$, in the same way.

Once the good parameters are chosen, for the first iteration, we increment the number of iterations, and we look for the good ones for the second iteration. Then we come back without decrementing the number of iterations so as to adjust the parameters $\alpha$ and $\beta$ for eventual improvement of the performance. Afterwards, we increment the number of iterations and repeat again the same process until a maximal number of iterations is reached.

# 5 RESULTS AND DISCUSSION

In this section, the performances of generalized parallel concatenated RS (GPCB-RS), BCH (GPCB-BCH) and BCH-RS (GPCB-BCH-RS) codes are evaluated. Transmission over the additive white Gaussian noise (AWGN) channel and binary antipodal modulation are used. We are interested in the information bit error rate (BER) for different signal to noise ratios per information bit ($E_b/N_0$) in dB. There are many parameters which affect the performance of GPCB codes when decoded with iterative decoder. Here we study the effect of the following parameters on the decoder performance: the number of decoding iterations, the component codes, interleaver size and patterns (see table 2)

TABLE 2 : THE PARAMETERS OF COMMUNICATION SYSTEM

| Parameter name | Value |
|---|---|
| Modulation | BPSK |
| Channel | AWGN |
| Interleaver pattern | Random interleaver (default value)<br>Diagonal interleaver<br>Cyclic interleaver<br>Block interleaver<br>Helical interleaver<br>Berrou's interleaver |
| $\alpha$ | 0.0, 0.25, 0.3, 0.4, 0.5, 0.55, 0.6, 0.65, 0.65, 0.7, 0.75, 0.80, 0.85, 0.9, 0.92, 0.95 |
| $\beta$ | 0.2, 0.25, 0.3, 0.35, 0.4, 0.45, 0.5, 0.55, 0.6, 0.65, 0.7, 0.75, 0.80, 0.85, 0.87, 0.9 |
| Component decoder | Chase-Pyndiah algorithm |
| Iterations number | 1 to 8 (default value) |
| Interleaver size | 1xk, 10xk, 100xk, 1000xk (k is the code dimension) |

## 5.1 GPCB-RS codes

Figure 3 shows the performance of the code GPCB-RS (73, 53), with M =100. This figure shows that the slope of curves and coding gain are improved by increasing the number of iterations.

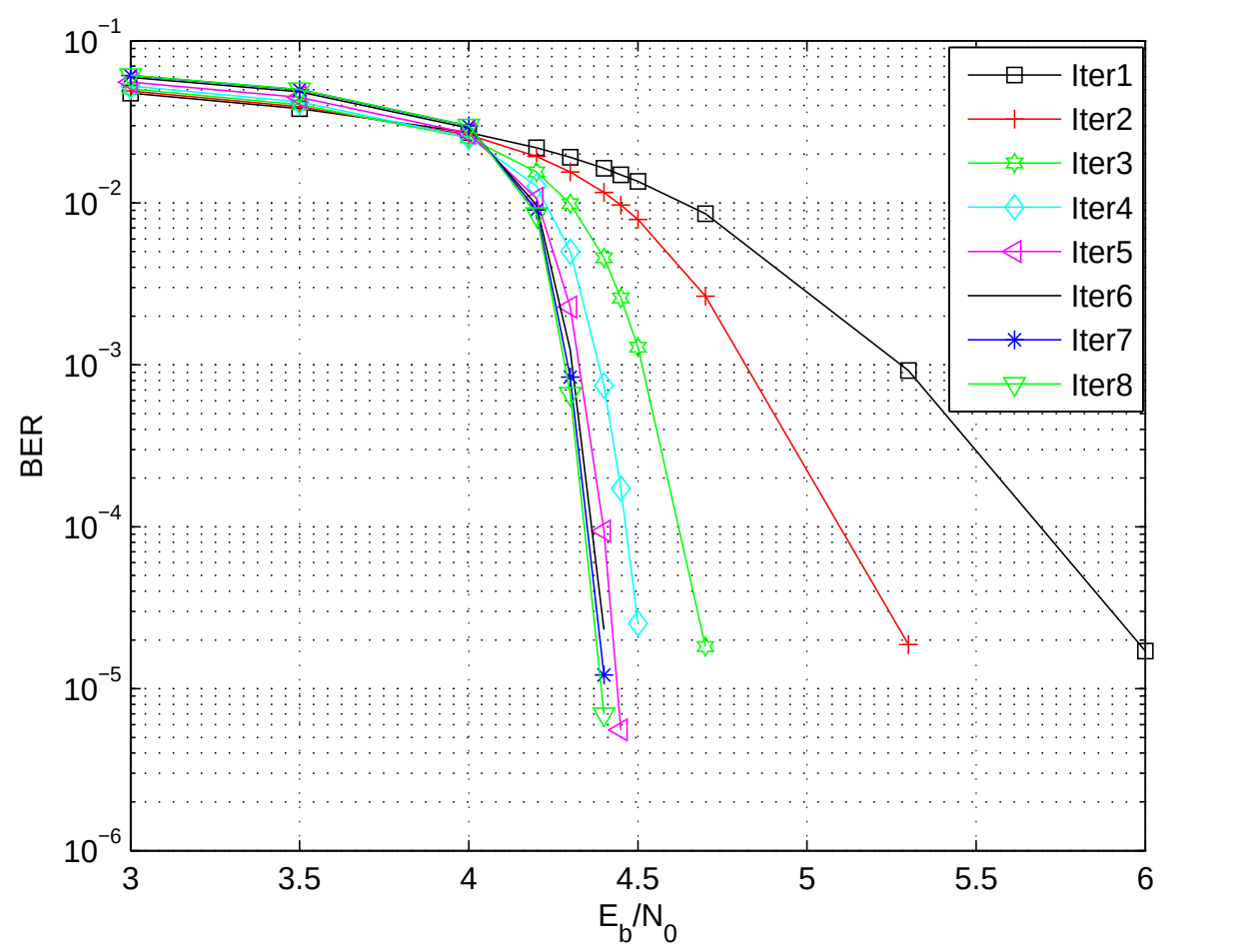

Figure 3 : Effect of iterations on iterative decoding of GPCB-RS(73, 53) Code, with M=100, over AWGN channel

At $10^{-5}$ about 1.65 dB coding gain can be obtained after 8 iterations. After the 8th iteration, the amelioration of the coding gain becomes negligible because of the steep slope of the BER curve. The turbo phenomenon is well established. In the rest of this paper, the curves of GPCB-RS codes are done with 8 iterations.

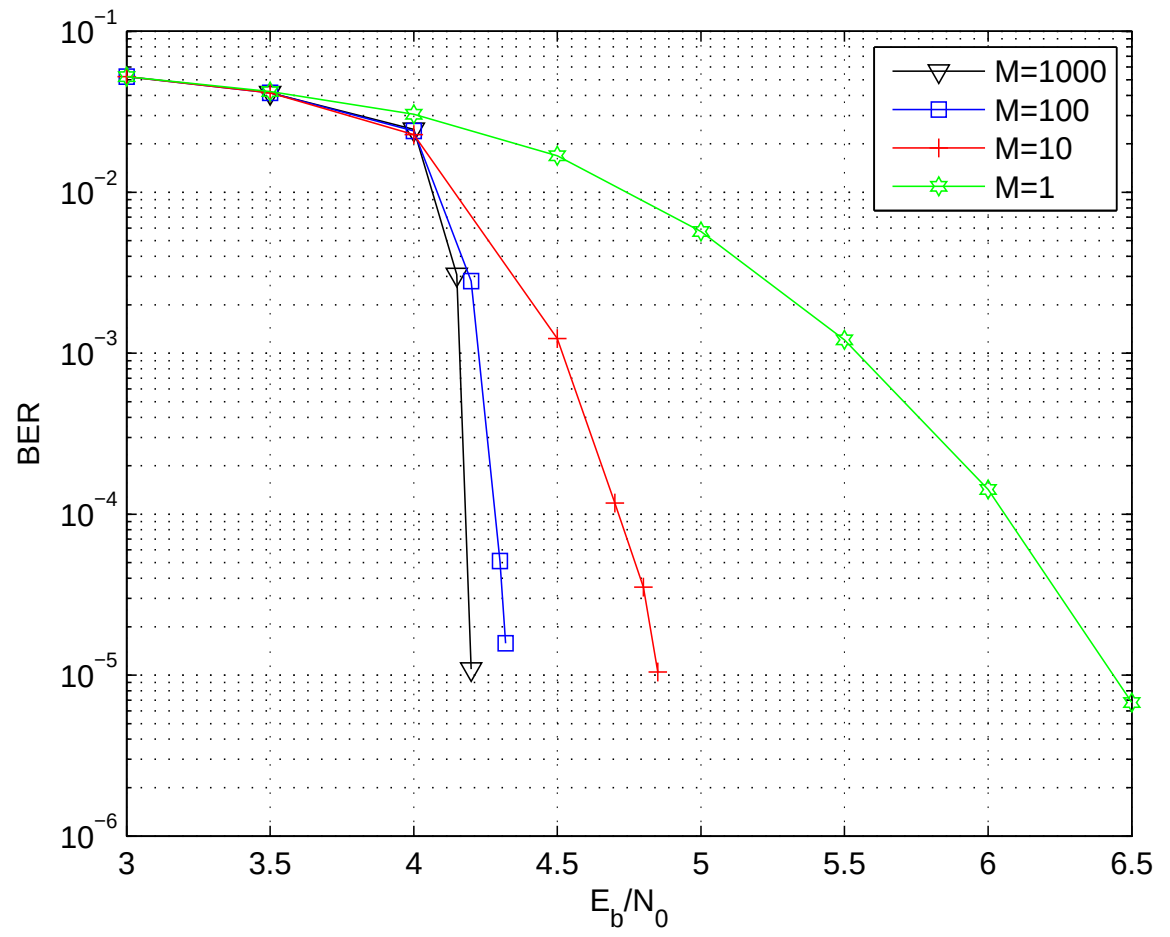

Figure 4: Effect of the parameter M on Iterative decoding of GPCB-RS (75, 51) code, over AWGN channel.

Figure 4 shows the BER versus SNR results of the GPCB-RS (75, 51) code with M varying from 1 to 1000. By increasing M from 1 to 100, about 2.1dB coding gain can be obtained at BER=$10^{-5}$ and little gain can be obtained by further increasing the parameter M.

Now, we consider the GPCB-RS (141,113) code. Its performance is shown in figure 5. For this code, the coding gain increase with M. At $10^{-5}$ about 1.6dB coding gain can be obtained by increasing M from 1 to 100. The amelioration becomes inconsiderable while the parameter M is greater than 100.

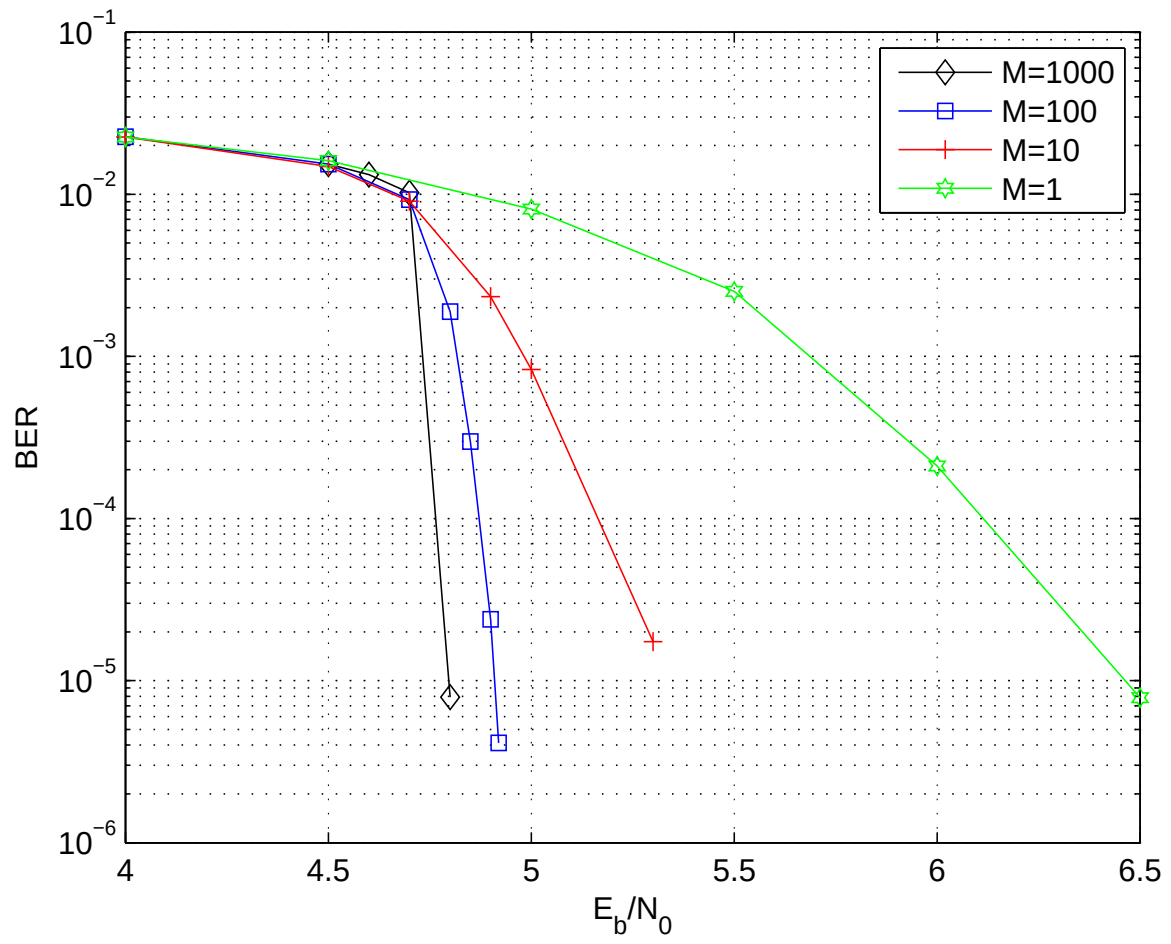

Figure 5: Effect of the parameter M on Iterative decoding of GPCB-RS (141, 113) code, over AWGN channel

The performance of the code GPCB-RS (271, 239) is shown in figure 6. For this code, the coding gain increase with M. At $10^{-5}$ about 1.0 dB coding gain can be obtained by increasing M from 1 to 100. The amelioration becomes inconsiderable while the parameter M is greater than 100.

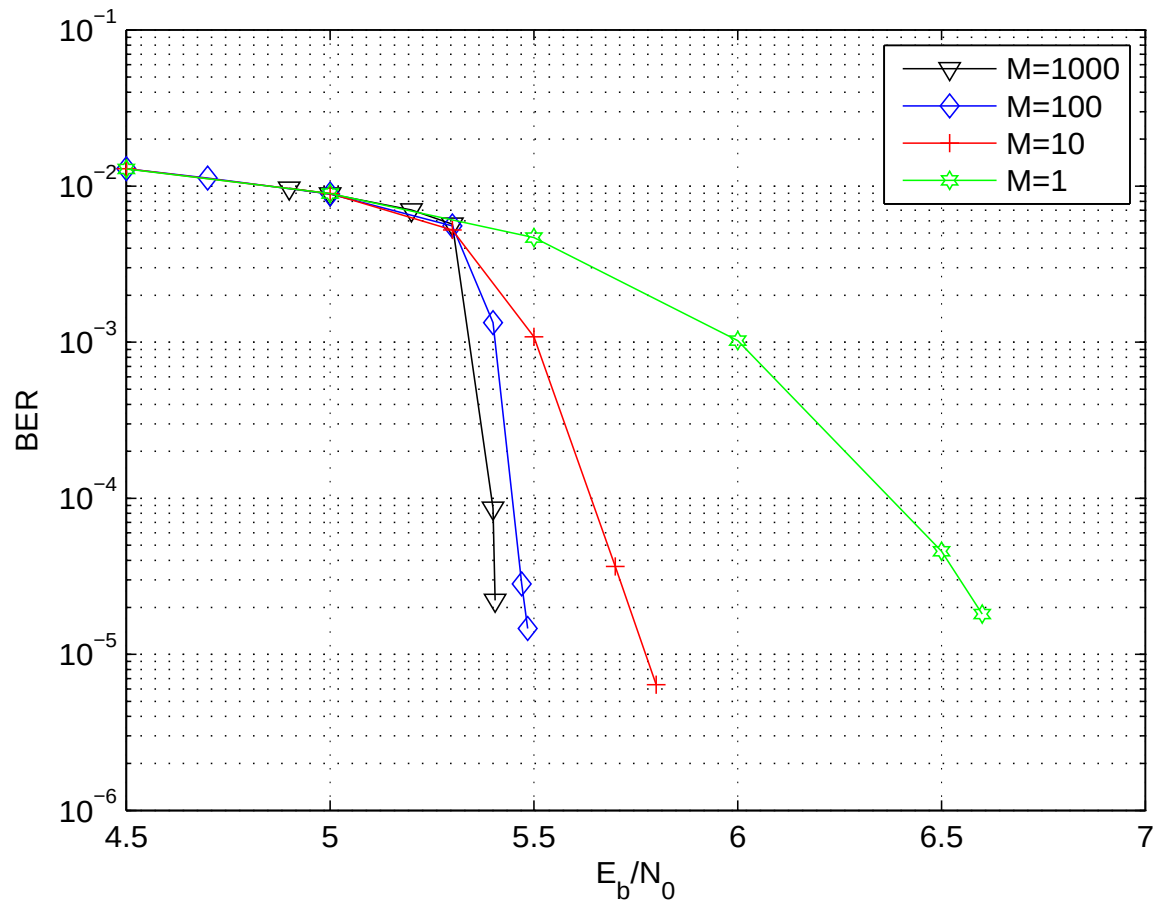

Figure 6: Effect of the parameter M on Iterative decoding of GPCB-RS (271, 239) code, over AWGN channel

To evaluate the performance of the parallel concatenated block codes, we compare the coding gain at the 8th iteration of the following codes GPCB-RS (69, 57), GPCB-RS (139, 115) and GPCB-RS (279, 231), with the same code rate 0.82 and the parameter M=100. The performance is shown in figure 7. From this figure, we observe that the performance becomes worse with increasing the length of the component code. The GPCB-RS (69, 57), GPCB-RS (139, 115) and GPCB-RS (279, 231) codes are respectively 2.27, 2.77 and 3.0dB away from their Shannon limits.

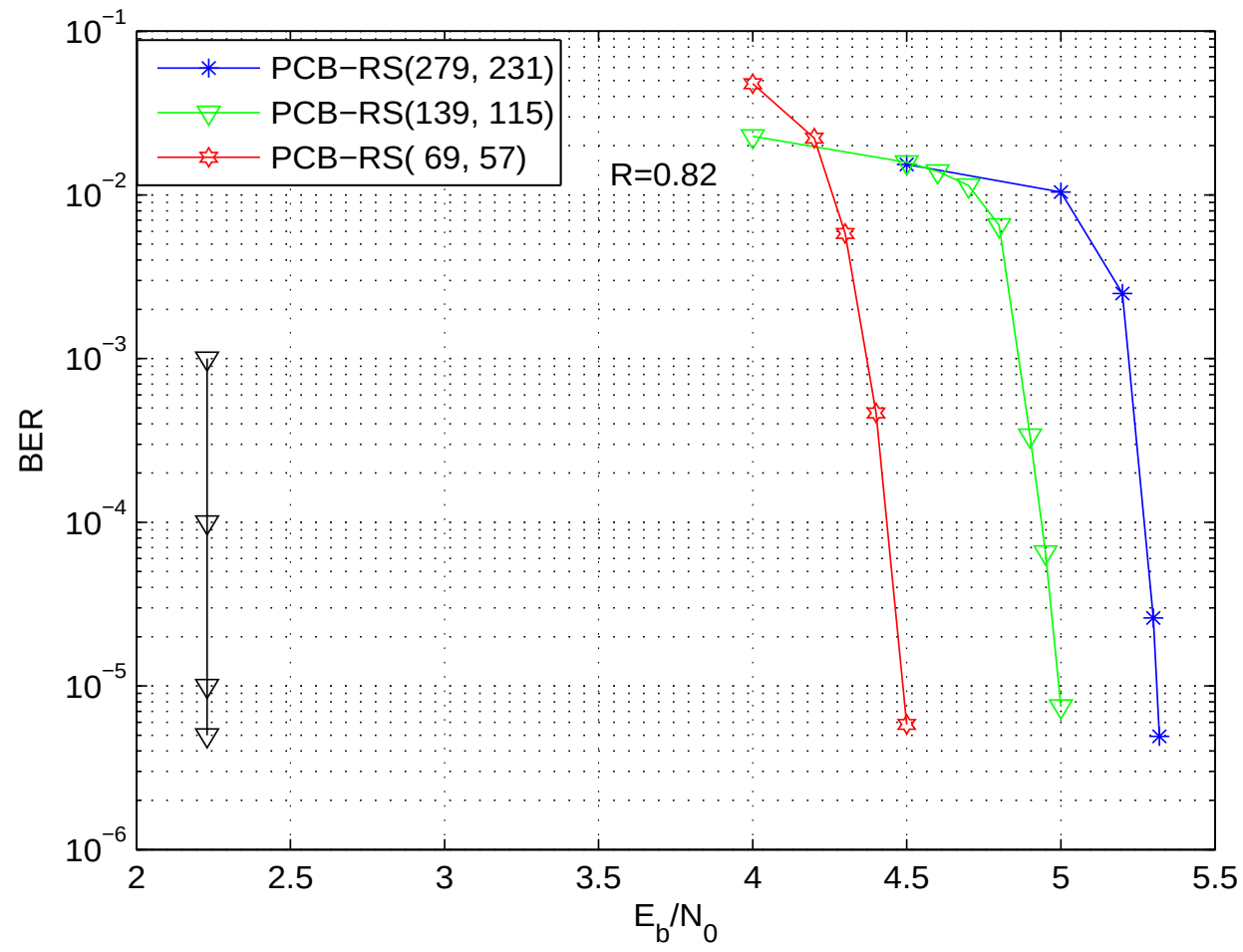

Figure 7: Performance evaluation of GPCB-RS codes, with M=100, over AWGN channel

To study the influence of the interleaver pattern on the performance of GPCB-RS codes, we have evaluated the BER of the GPCB-RS (73, 53) code using various interleaver patterns such as diagonal, cyclic, block, helical and random interleaver with the parameter M=1000. The figure 8 shows the result. We observe that the best interleaver is the random interleaver, followed by helical, Block and Berrou's interleaver then diagonal and cyclic interleaver.

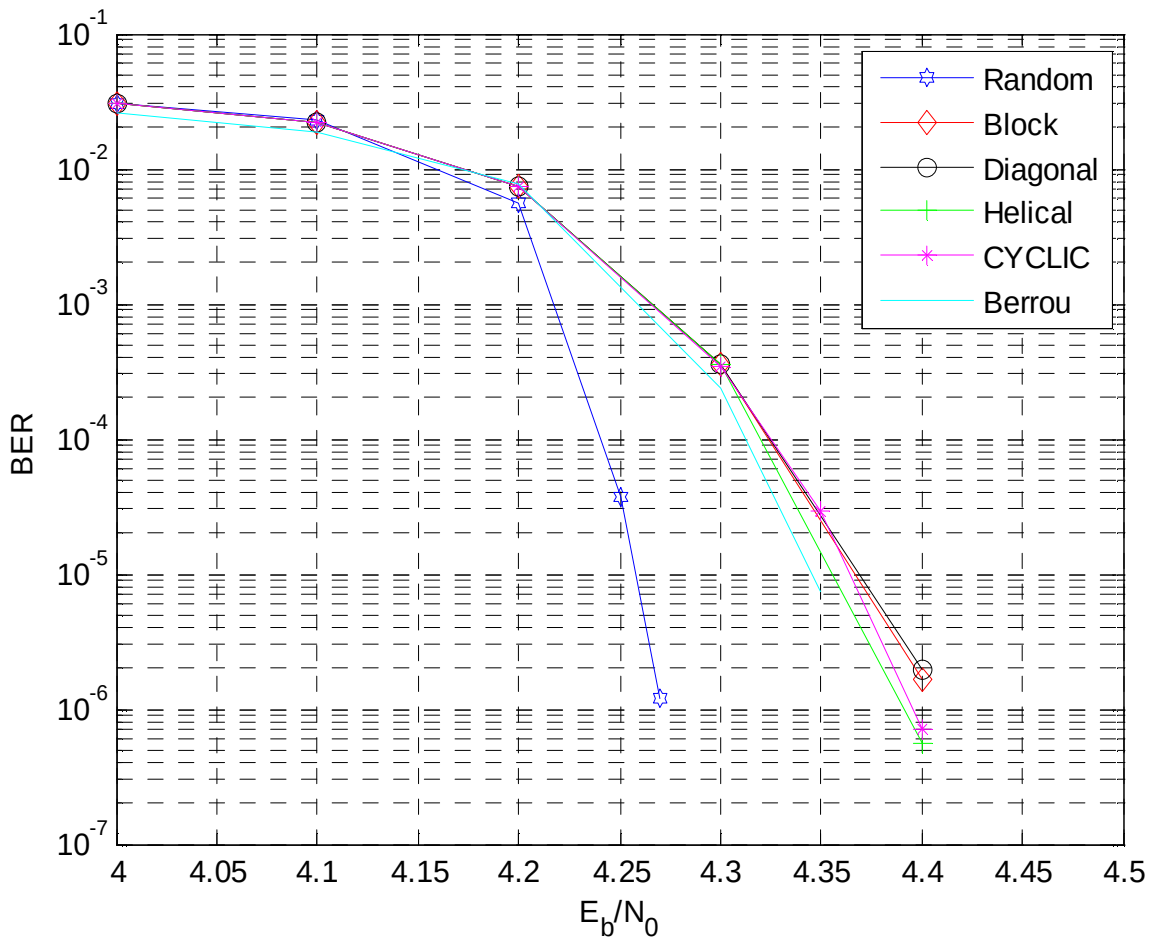

Figure 8: Interleaver structure effect on iterative decoding of GPCB-RS(73, 53) code, with M=1000, over AWGN channel

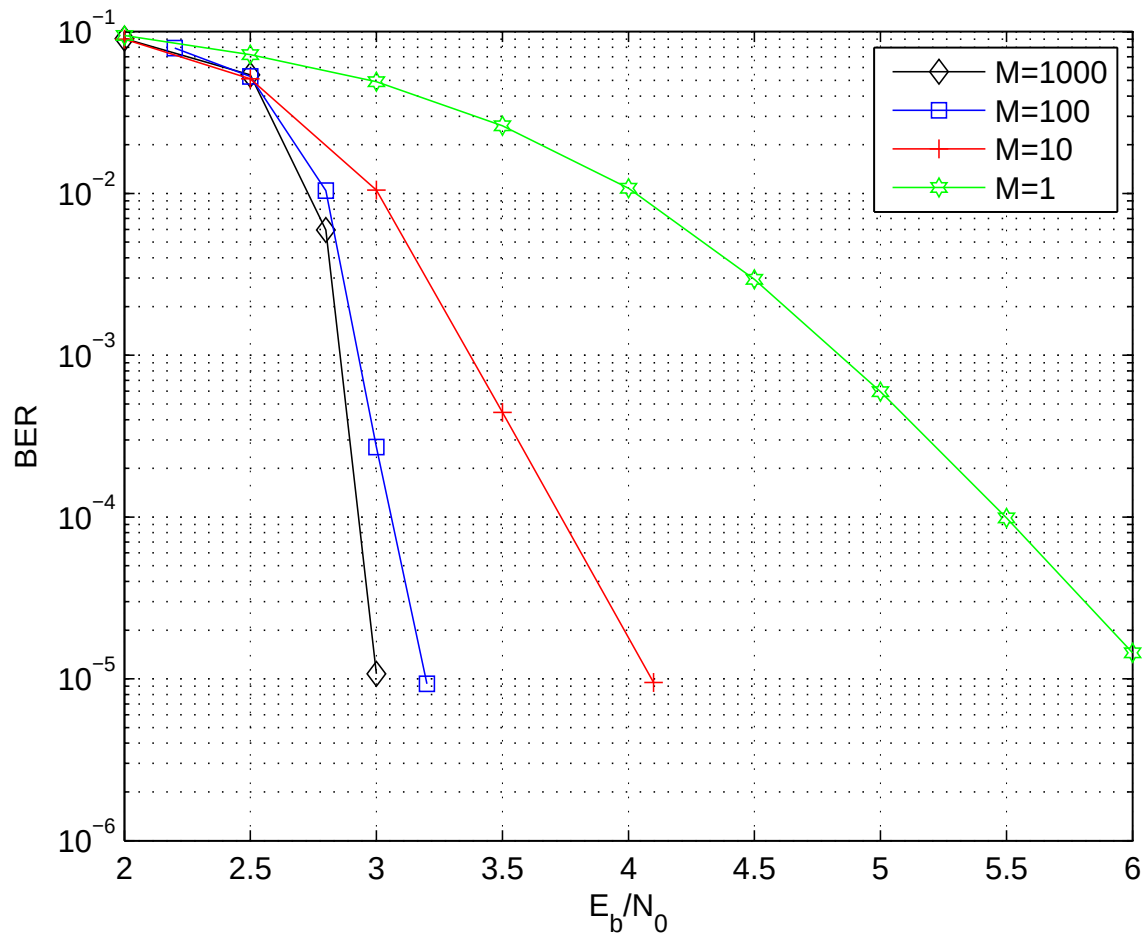

Figure 10: Effect of the parameter M on Iterative decoding of GPCB-BCH (148, 106) code, over AWGN channel

## 5.2 GPCB-BCH codes

Figure 9 shows the performance of the code GPCB-BCH (148, 106), with M =100. This figure shows that the coding gain is improved by increasing the number of iterations. At $10^{-5}$ about 2.5 dB coding gain can be obtained after 7 iterations. After the 7th iteration, the amelioration of the coding gain becomes negligible. The turbo phenomenon is well established. In the rest of this paper, for GPCB-BCH codes, the curves are done with 7 iterations.

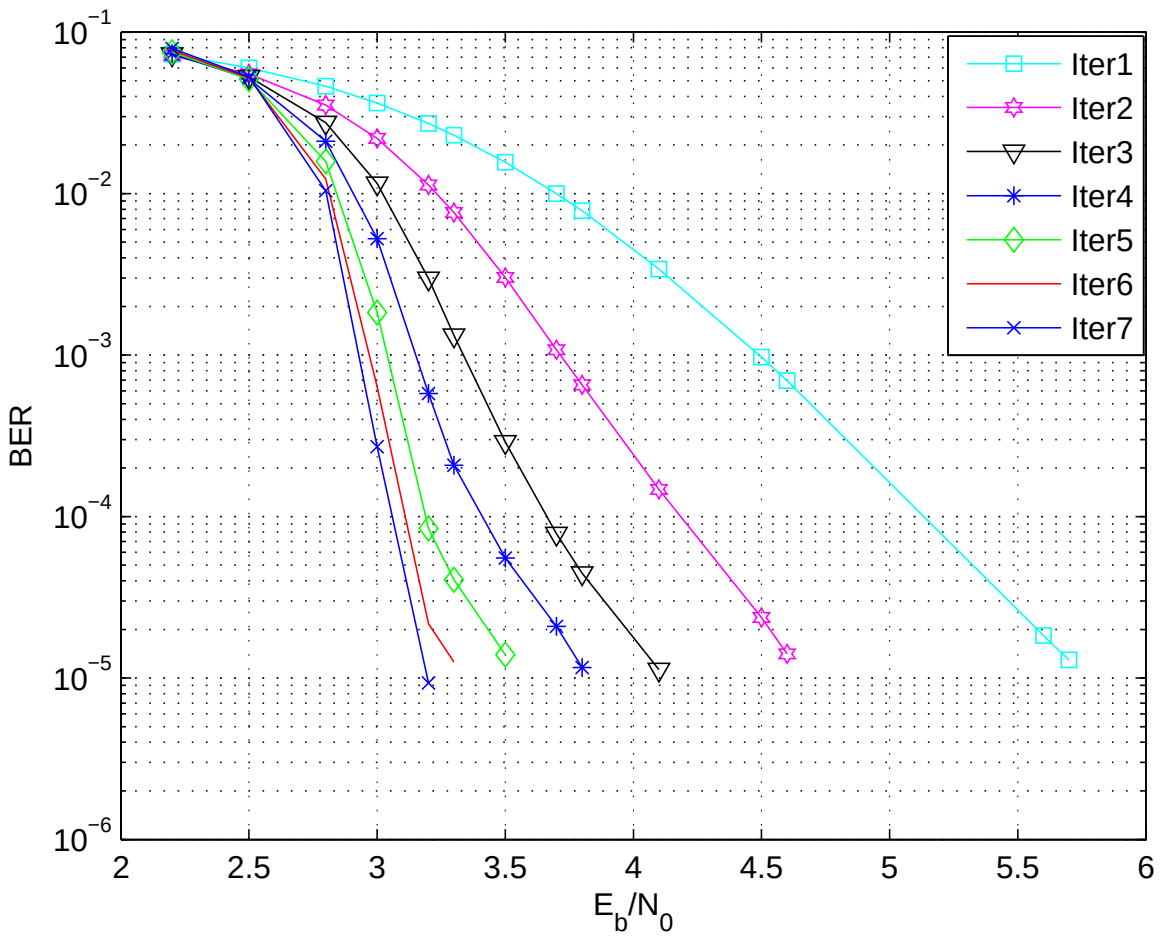

Figure 9: Effect of iterations on iterative decoding of GPCB-BCH(148, 106) Code, over AWGN channel

In order to study the effect of the parameter M, we consider the GPCB-RS (148,106) code. Its performance is shown in figure 10. For this code, the coding gain increase with M. At $10^{-5}$ about 3 dB coding gain can be obtained by increasing M from 1 to 100. The amelioration becomes inconsiderable while the parameter M is greater than 100.

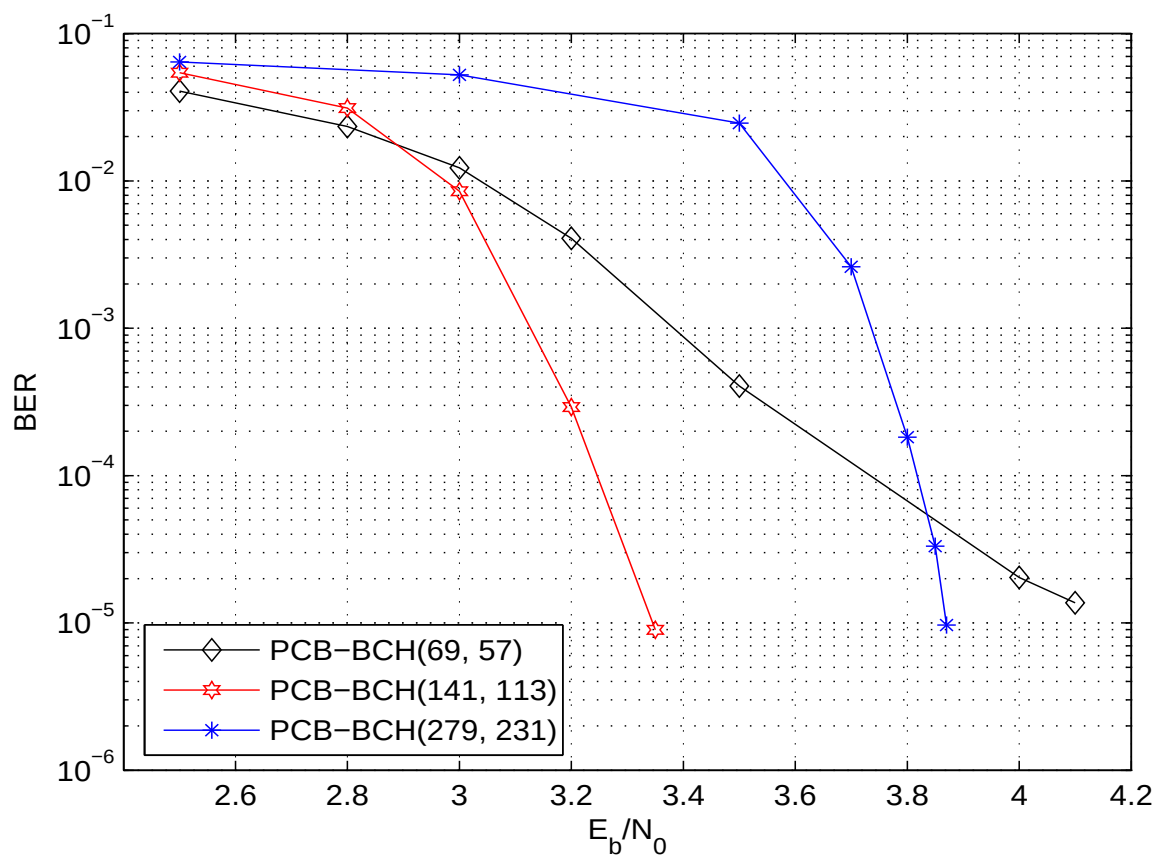

Figure 11: Performance evaluation of GPCB-BCH codes, with M=100, over AWGN channel

To evaluate the performance of the parallel concatenated block codes, we compare the coding gain at the 7th iteration of the following codes GPCB-BCH (69, 57), GPCB-BCH (141, 113) and GPCB- BCH (279, 231), with almost the same code rate 0.82 and the parameter M=100. The performance is shown in figure 11. From this figure, we observe that the performance of the code GPCB-BCH (69, 57) is worse than the one of GPCB-BCH (279, 231) codes. The later code in his role is worse than GPCB-BCH (141, 113). The GPCB-BCH (69, 57), GPCB-BCH (141, 113) and GPCB-BCH (279, 231) codes are respectively 1.9, 1.3 and 1.6 dB away from their Shannon limits.

To study the influence of the interleaver pattern on the PCB codes performance, we have evaluated the BER versus $E_b/N_0$ of the PCB-BCH (148, 106) code using different interleaver structures such as diagonal, cyclic, block and random interleaver with parameter M=1000. The figure 12 shows the performance results. We observe that the random interleaver is little good than block, diagonal and cyclic ones.

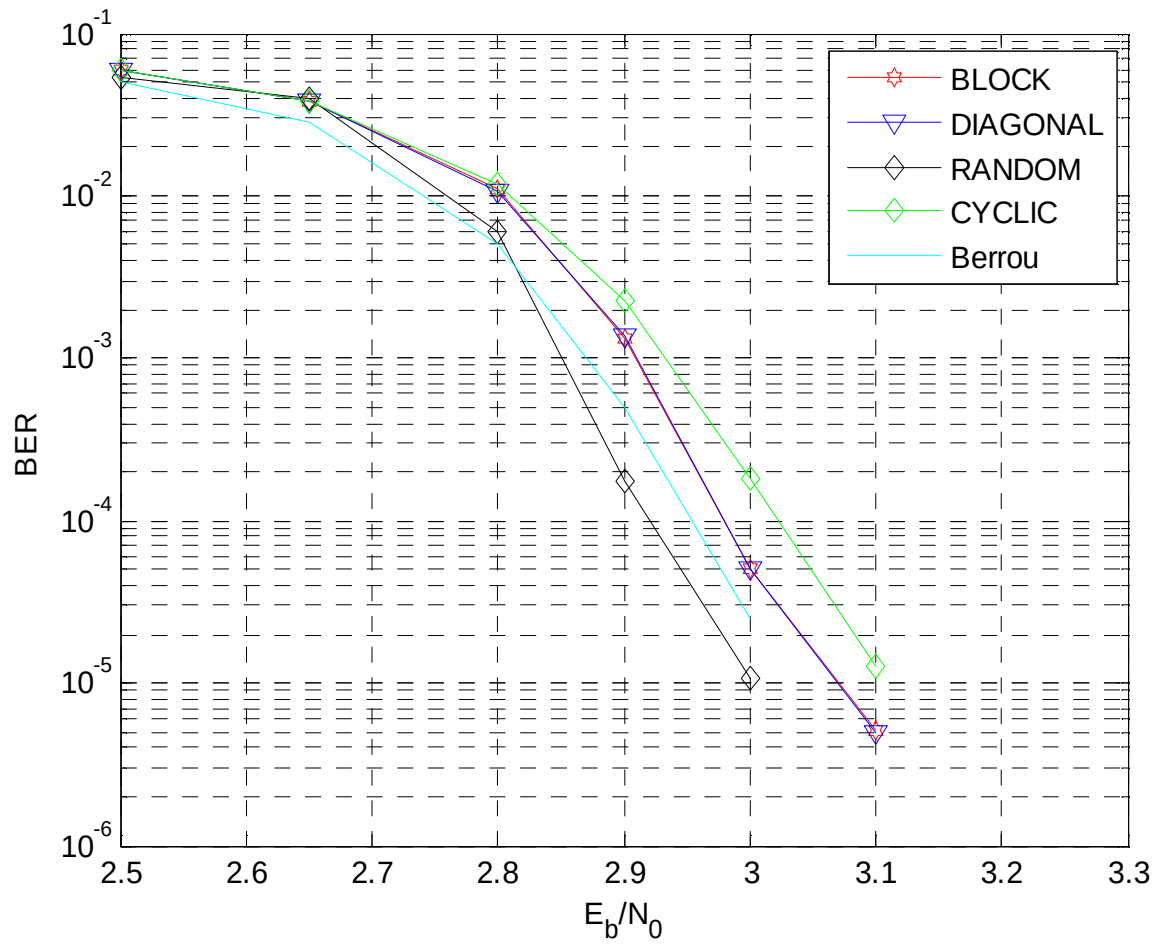

Figure 12: Interleaver structure effect on Iterative decoding of the GPCB-BCH(148, 106) code, with M=1000, over AWGN channel

## 5.3 GPCB-BCH-RS codes

To study the iteration effects of the GPCB-BCH-RS codes, We plot the performance of GPCB-BCH-RS(141, 113) code, with M =100, in the figure 13. The later shows that coding gain is improved by increasing the number of iterations. At $10^{-5}$ about 1.2 dB coding gain can be obtained at 8th iterations. After the 8th iteration, the amelioration of the coding gain becomes negligible. The turbo phenomenon is well established. In the remaining part of this paper, for GPCB-BCH-RS codes, the curves are done with 8 iterations.

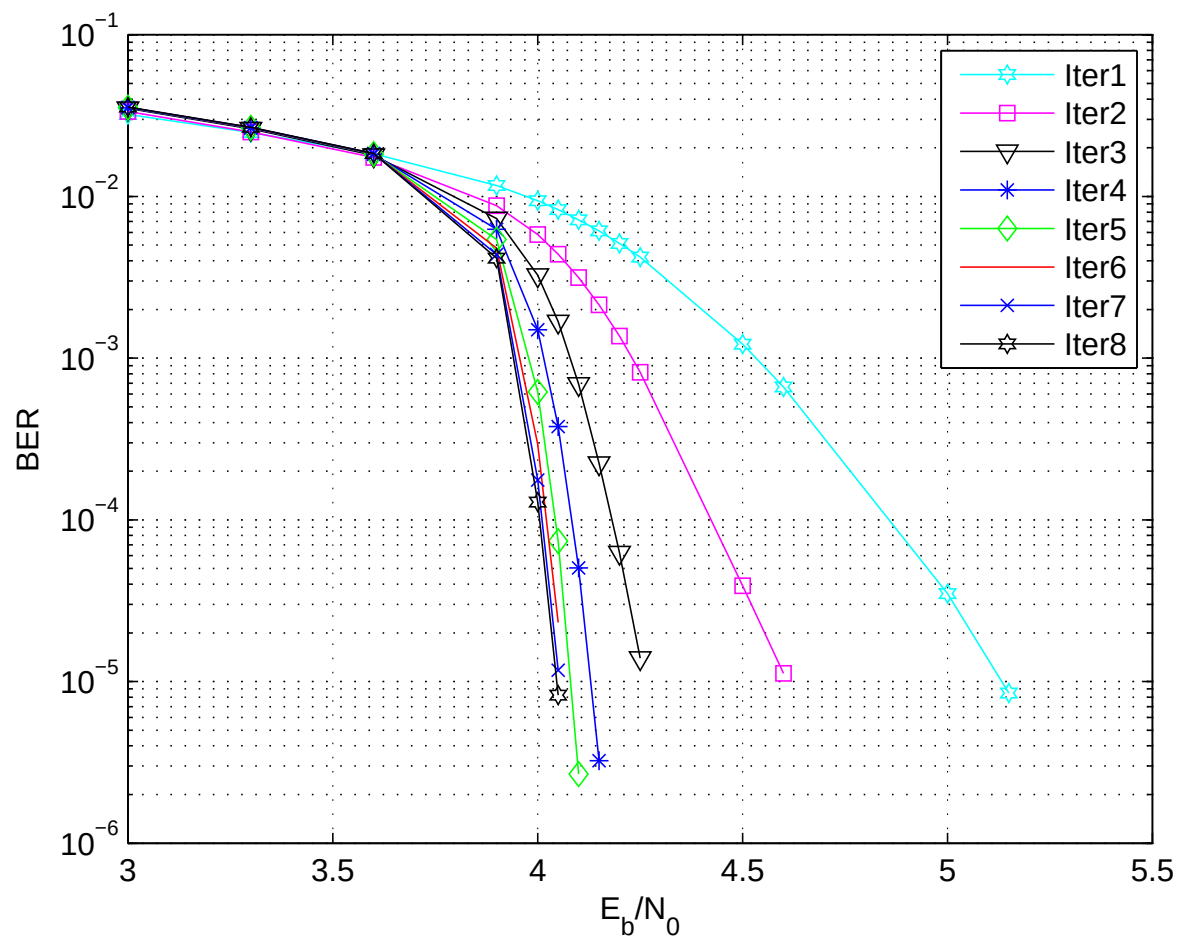

Figure 13: Effect of iterations on iterative decoding of GPCB-BCH-RS(141, 113) Code, with M=100, over AWGN channel

We have evaluated the effect of the parameter M on the performance GPCB-BCH-RS codes. The figure 7 depict BER versus $E_b/N_0$ of the GPCB-BCH-RS(141, 113). The coding gain increases with the M. At $10^{-5}$ about 1.3 dB coding gain can be obtained by increasing M from 1 to 1000. The amelioration becomes inconsiderable while the parameter M is greater than 100.

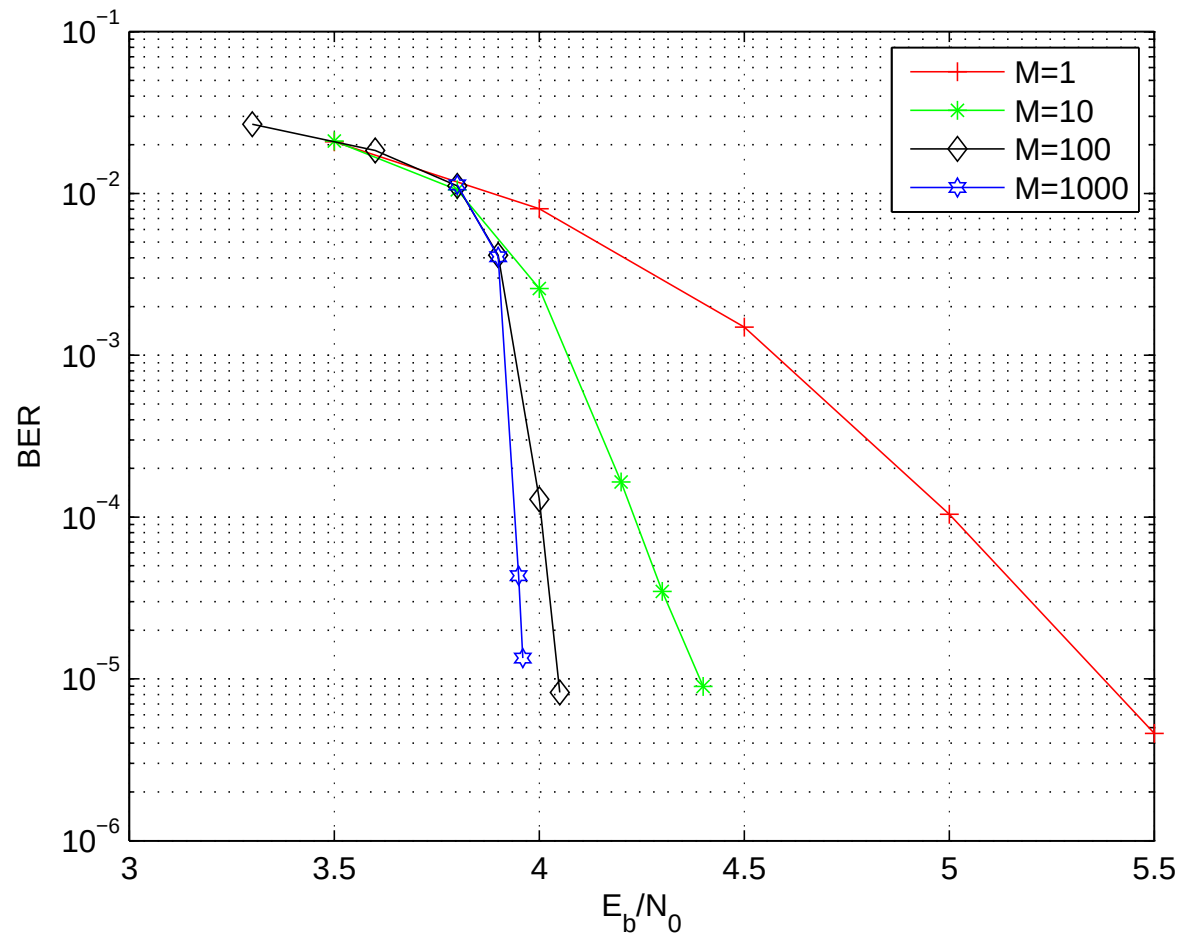

Figure 14: Effect of the parameter M on Iterative decoding of GPCB-BCH-RS (141, 113) code, over AWGN channel

To study the influence of the interleaver pattern on the PCB-BCH-RS codes performance, we have evaluated the BER versus $E_b/N_0$ of the PCB-BCH-RS (141, 113) code using different interleaver structures such as diagonal, cyclic, block and random interleaver with parameter M=100. The figure 15 shows the performance results. We observe that these interleavers are comparable for this code.

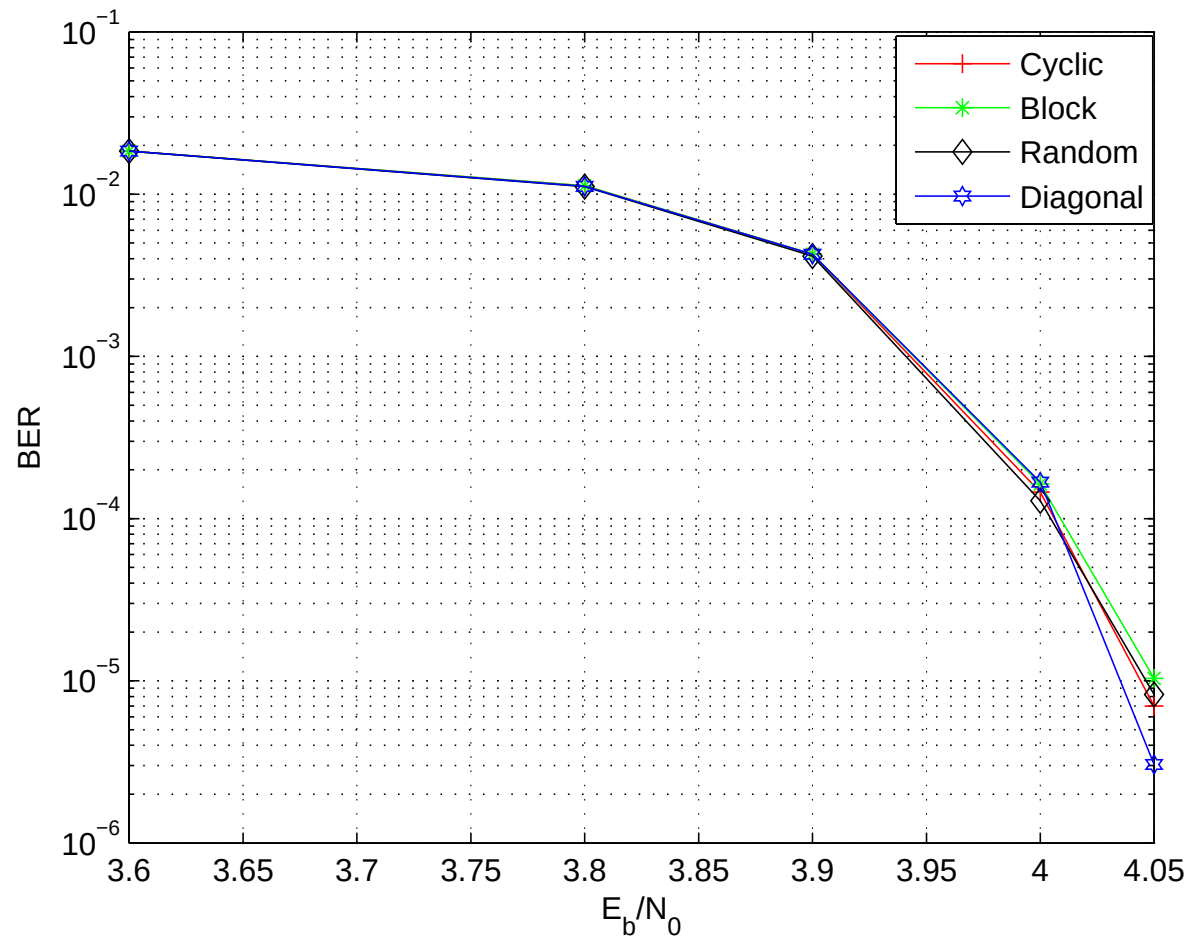

Figure 15: Interleaver structure effect on Iterative decoding of the GPCB-BCH-RS(141, 113) code, with M=100, over AWGN channel

To evaluate the performance of the GPCB-BCH-RS codes, we compare the coding gain at the 8th iteration of the following codes GPCB-BCH-RS (69, 57), GPCB-BCH-RS (141, 113) and GPCB-BCH-RS (279, 231), with almost the same code rate 0.82 and the parameter M=100. The performance is shown in figure 16. From this figure, we observe that the performance of the code GPCB-BCH (69, 57) is worse than the one of GPCB-BCH (279, 231) codes. The later code in his role is worse than GPCB-BCH (141, 113).

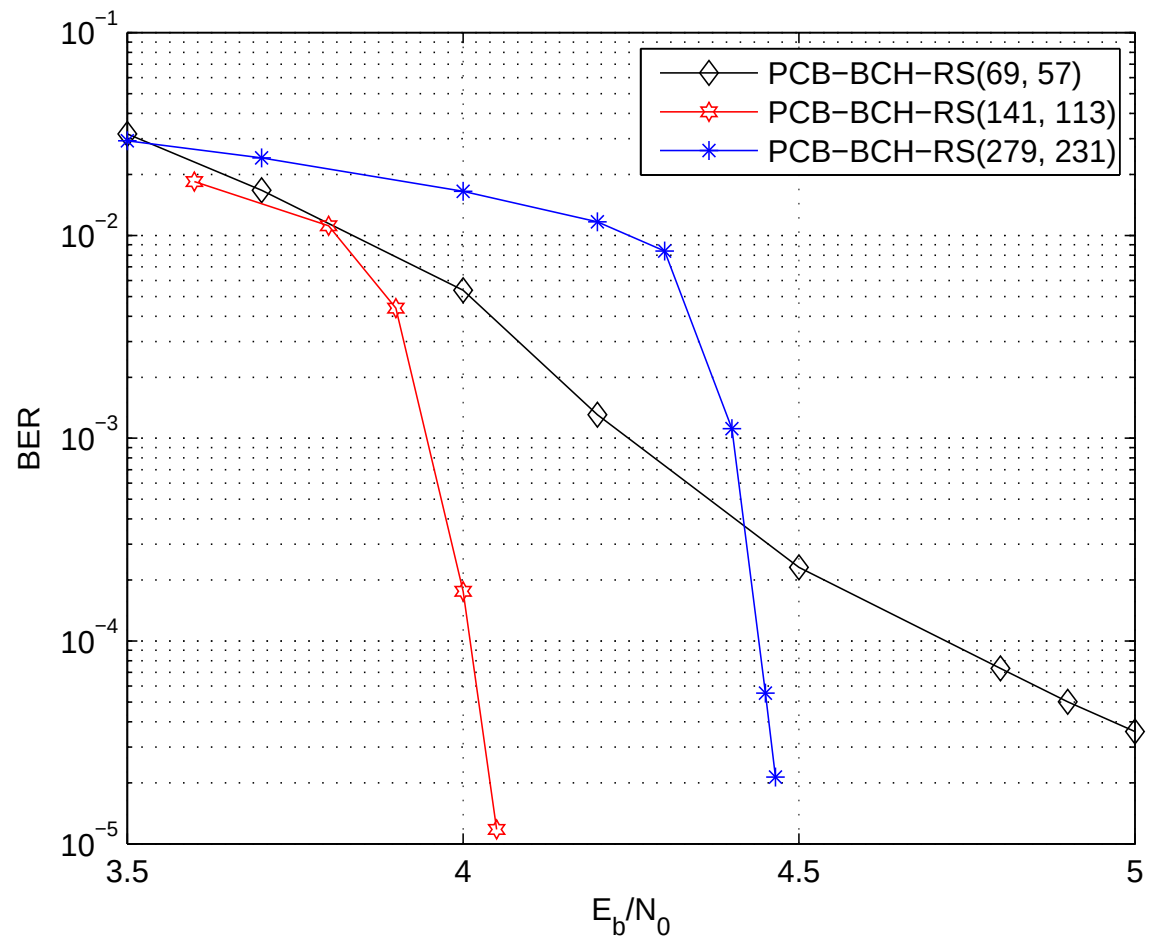

Figure 16: Performance evaluation of GPCB-BCH-RS codes, with M=100, over AWGN channel

## 6 COMPARISON BETWEEN GPCB-RS, GPCB-BCH AND GPCB-RS-BCH CODES.

To compare between GPCB-BCH, GPCB-RS, and GPCB-BCH-RS codes we have evaluated the performances of three codes GPCB-BCH(141, 113), GPCB-BCH-RS(141, 113) and GPCB-RS(141, 113) with the same rate, 0.80. The results are shown in the figure 17. According to this figure, we observe that GPCB-BCH code outperform GPCB-BCH-RS by 0.6dB, and GPCB-BCH-RS code outperform GPCB-RS. So, the GPCB-BCH code is the best one.

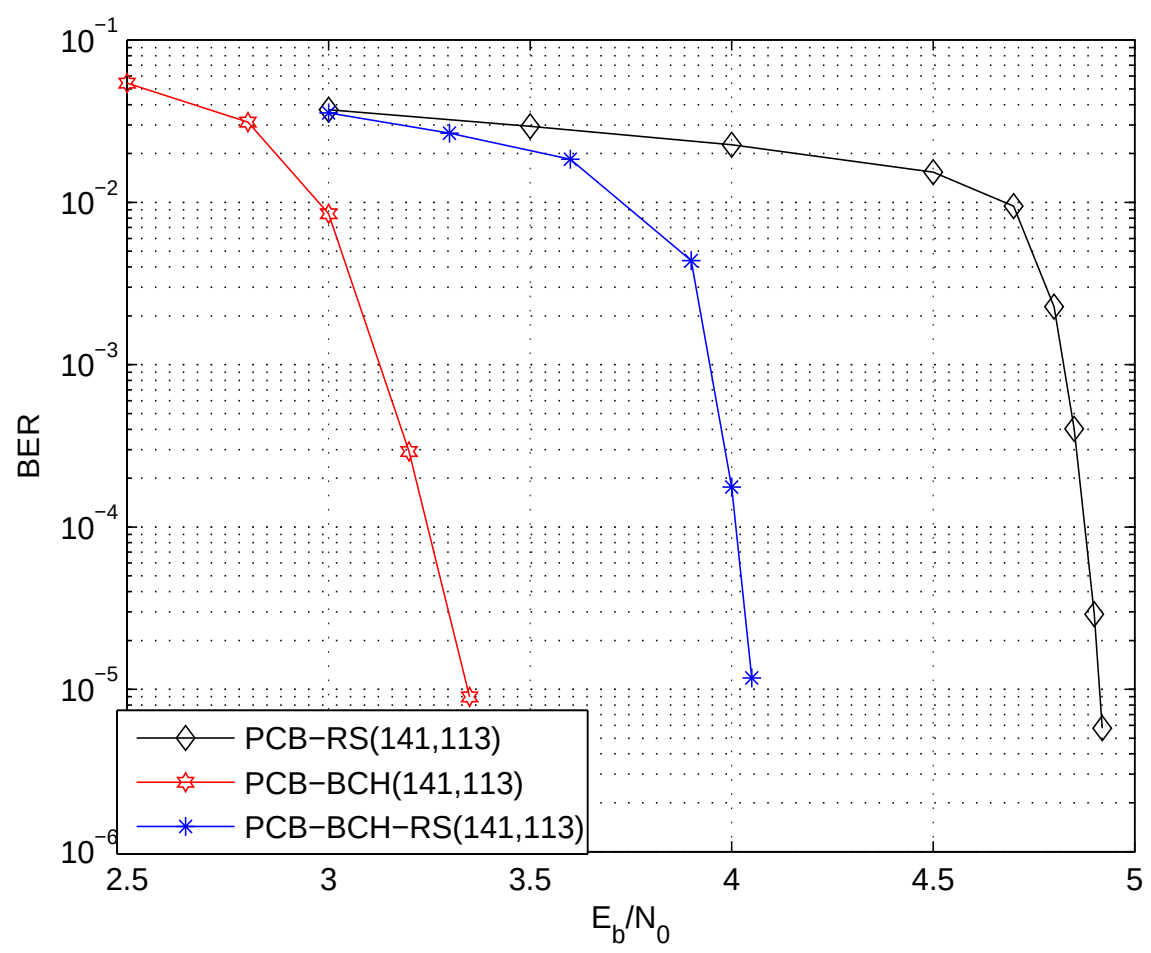

Figure 17: Performances comparison between GPCB-RS, GPCB-BCH-RS and GPCB-BCH codes, with M=100, over AWGN channel.

## 7 CONCLUSION

In this paper, a generalization of parallel concatenated block (GPCB) codes based on RS and BCH codes is presented. We demonstrated the relevance of the Chase-Pyndiah SISO algorithm for the decoding of GPCB codes. Moreover, we investigated the effects of various factors on the performances of these codes such as: component codes, the number of iterations, interleaver sizes and patterns using simulations. The simulation results shows that the slope of curves and coding gain are improved by increasing the number of iterations and/or the interleaver size (the parameter M). The comparison between GPCB-BCH, GPCB-BCH-RS and GPCB-RS codes shows that the performance of the GPCB-BCH codes is the best one, and GPCB-BCH-RS codes fall in between the two other GPCB codes.
The performance of GPCB-RS and GPCB-BCH-RS codes becomes worse when increasing the length of the component code. To overcome this problem, we propose to develop a new decoder for GPCB codes. The obtained results by applying the above construction and decoding codes look very promising and open new perspectives.

## APPENDIX

The number of the test-sequences, $Y^l$, used by Chase-Pyndiah decoding is 18. Let $I_1$, $I_2$, $I_3$, $I_4$ and $I_5$ denote the positions of the five least reliable symbols at the input of the component decoder. These five positions are classed in increasing reliability order. The first test-sequence is, $Y^0$, the hard decision of the input of the Chase-Pyndiah decoder, the other test-sequences are given below. Between brackets are the non null positions for each sequence.

$Y^1$ $(I_1)$
$Y^2$ $(I_2)$
$Y^3$ $(I_1, I_2)$
$Y^4$ $(I_3)$
$Y^5$ $(I_1, I_3)$
$Y^6$ $(I_4)$
$Y^7$ $(I_2, I_3)$
$Y^8$ $(I_1, I_4)$
$Y^9$ $(I_1, I_2, I_3)$
$Y^{10}$ $(I_1, I_5)$
$Y^{11}$ $(I_2, I_3, I_4)$
$Y^{12}$ $(I_1, I_2, I_3, I_4)$
$Y^{13}$ $(I_1, I_3, I_5)$
$Y^{14}$ $(I_1, I_2, I_4, I_5)$
$Y^{15}$ $(I_1, I_3, I_4, I_5)$
$Y^{16}$ $(I_2, I_3, I_4, I_5)$
$Y^{17}$ $(I_1, I_2, I_3, I_4, I_5)$

## REFERENCES

[1] C Berrou, A. Glavieux, P. Thitimajshima, "Near Shannon limit error correcting coding and decoding: TurboCodes", IEEE International Conference on Communication ICC93, vol. 2/3, May 1993.

[2] R. Pyndiah, A. Glavieux, A. Picart, S. Jacq, "Near optimum decoding of product codes", GLOBECOM94, November 1994.

[3] D. Chase, "Class of algorithms for decoding block codes with channel measurement information", IEEE Trans. Information theory, Vol. 13, pp. 170-182, Jan.1972

[4] S. Benedetto and G. Montorsi, "Average Performance of Parallel Concatenated Block Codes," Electronics Letters, vol. 31, no. 3, pp. 156-158, February 1995.

[5] P. Adde, R. Pyndiah, O. Raoul, "Performance and complexity of block turbo decoder circuits", Proceedings of the Third IEEE International Conference Electronics, Circuits, and Systems, ICECS , pp.172 – 175, Oct 1996.

[6] J. Nilsson, R Kotter "Iterative decoding of product code constructions" Proceedings of ISIT-94, pp.1059-1064, Sydney, Australia, November 1994.

[7] M.Belkasmi, A.Farchane "Iterative decoding of parallel concatenated block codes" , Proceedings of the International Conference on Computer and Communication Engineering 2008 May 13-15, 2008 Kuala Lumpur, Malaysia

[8] Sergio Benedetto, and Guido Montorsi, "Unveiling Turbo Codes: Some Results on Parallel Concatenated Coding Schemes", IEEE Transactions on Information Theory, Vol. 42, No.2, March 1996.

[9] C. Berrou, A. Glavieux, "Near Optimum Error Correcting Coding and Decoding": Turbo Codes, IEEE Transactions on Communications, vol. 44, October 1996.

[10] Consultative Committee for Space Data Systems "Telemetry Channel Coding" Blue Book 2002, 101.0-B-6.

[11] Aitsab and R. Pyndiah "performance of concatenated Reed-Solomon/convolutional codes with iterative decoding", IEEE GLOBCOM'97, pp.934-938, Phoenix, Nov.1997.

**A. Farchane** received his license in Computer Science and Engineering in June-2001 and Master in Computer Science and telecommunication from University of Mohammed V - Agdal, Rabat, Morocco in 2003. Currently he is doing his PhD in Computer Science and Engineering at ENSIAS (Ecole Nationale Supérieure d'Informatique et d'Analyse des Systèmes), Rabat, Morocco. His areas of interest are Information and Coding Theory.

**M. Belkasmi** is a professor at ENSIAS (Ecole Nationale Supérieure d'Informatique et d'Analyse des Systèmes,Rabat); head of Telecom and Embedded Systems Team at SIME Lab.He had PhD at Toulouse University in 1991(France). His current research interests include mobile and wireless communications, interconnexions for 3G and 4G, and Information and Coding Theory.

**S.Nouh,** received his license in Computer Science and Engineering in June-2003 and Master in information processing techniques from University of Hassan II, Mohammedia (Morocco) in 2005. Currently he is doing his PhD in Computer Science and Engineering at ENSIAS (Ecole Nationale Supérieure d'Informatique et d'Analyse des Systèmes), Rabat, Morocco. His areas of interest are Information and Coding Theory.